 \documentclass[final,5p,times,twocolumn]{elsarticle}

\usepackage{amssymb}
\usepackage{lipsum}
\usepackage{algorithm}
\usepackage{algpseudocode}
\usepackage{hyperref}
\usepackage{float}
\usepackage{multicol}
\usepackage{algpseudocode}
\usepackage{algorithm}
\usepackage{longtable}
\usepackage{tikz}
\def\checkmark{\tikz\fill[scale=0.4](0,.35) -- (.25,0) -- (1,.7) -- (.25,.15) -- cycle;}

\journal{Journal of Information Security and Application}

\begin{document}

\begin{frontmatter}



\title{Enforcing Data Geolocation Policies in Public Clouds using Trusted Computing}

\author{Syed Zair Abbas}

\affiliation{organization={FAST National University of Computer and Emerging Sciences},
            addressline={}, 
            city={Islamabad},
            postcode={}, 
            state={},
            country={Pakistan}}

 \author{Mudassar Aslam}
 
 \affiliation{organization={FAST National University of Computer and Emerging Sciences},
             addressline={},
             city={Islamabad},
            postcode={}, 
            state={},
            country={Pakistan}}


\begin{abstract}
With the advancement in technology, Cloud computing always amazes the world with revolutionizing solutions that automate and simplify complex computational tasks. The advantages like no maintenance cost, accessibility, data backup, pay-per-use models, unlimited storage, and processing power encourage individuals and businesses to migrate their workload to the cloud. Despite the numerous advantages of cloud computing, the geolocation of data in the cloud environment is a massive concern, which relates to the performance and government legislation that will be applied to data. The unclarity of data geolocation can cause compliance concerns. In this work, we have presented a technique that will allow users to restrict the geolocation of their data in the cloud environment. We have used trusted computing mechanisms to attest the host and its geolocation remotely. With this model, the user will upload the data whose decryption key will be shared with a third-party attestation server only. The decryption key will be sealed to the TPM of the host after successful attestation guaranteeing the authorized geolocation and platform state.
\end{abstract}




\end{frontmatter}


\section{Introduction}
\label{}
Owing to the impact of cloud computing, more users are acquiring cloud services from cloud service providers who are expected to ensure the security, privacy, and availability of data. According to the National Institute of Standards and Technology (NIST), cloud computing allows us to acquire on-demand computing resources like storage, server, and software which can be provisioned or released without any extensive effort. Users and organizations are moving towards cloud solutions because now they no longer have to pay for maintenance of hardware, instead, they only have to deploy/upload their application/data on applications provided by cloud service providers. Customers can access their respective services from anywhere. They only have to pay for the resources they are consuming. The resources can be dynamically scaled out depending on customer demand. 

Cloud Service Providers (CSP) offer storage as a service. The users can upload and access their data from anywhere. The advantages of cloud storage are fast data access/retrieval, unlimited storage, and low cost because users only have to pay for the storage acquired by their data \cite{Yang2020, rashmi2020rdpc}. Users and organizations are moving towards cloud solutions because they are free from hardware maintenance costs. They only have to deploy/upload their application/data on platforms by cloud service providers. The data on the cloud can be accessed quickly as it is stored on multiple servers. In case of disaster when companies might lose the data or its backup which are stored locally, the cloud offers data redundancy by storing the backups in multiple availability zones. This redundancy also assists in a situation when data might be lost because of hardware failure. Users can acquire limitless storage virtually on cloud platforms \cite{Obrutsky2016, liu2018overview}.

The data is considered an asset. It is a topmost concern for individuals and organizations. In the data lifecycle which includes the creation, storage, secure usage, and removal of data, every CSP is expected to look after all these phases \cite{Basu2018}. When users or organizations outsource their data to the cloud, they lose physical control over their data. The Service Level Agreement (SLA) is a document containing the terms and conditions of the contract between the customers and users\cite{mirobi2015service}. The SLA can be used by the customers to define the geolocation where data should be stored and other security parameters that can be applied to ensure the availability, privacy, and security of data in a cloud environment\cite{dash2014service, ahmed2019brief}. The region in which data is stored or being processed has importance because of multiple factors such as performance which will be degraded if the data is stored far away from the region where it is being accessed by users, security, and privacy because few regions have specific laws for privacy protection, legal compliance can result in conflict with data sovereignty if the outsourced data is against the rules and regulations of the country or state where it is been stored \cite{Singi2020, gunawi2016does}. Further, a small misconfiguration can lead to unauthorized access to data and can lead to compliance issues\cite{achar2022data}. Generally, it is challenging for the users of the cloud to verify whether the data on the cloud is stored in the regions that are specified in SLA or not. The reasons can be: 
1) the users who acquire cloud services just need to upload their data on the platform provided by CSP without the knowledge of underlying infrastructure; 2) the CSP can move the data for workload balancing, hardware failure, or upgradation while maintaining the availability. The distributed nature of the cloud makes it more difficult for users to determine the geolocation \cite{Tabrizchi2020}. The worst thing is that the cloud users cannot even detect such data placement in unauthorized regions by the CSP. This highlights the need for a solution to enforce the access and confidentiality of data\cite{qi2019crypt, qi2016scalable}.

There are mainly two approaches to determining the geolocation of a data center: landmark-based and hardware-based \cite{Irain2017}. First, the landmark-based can be further classified into two groups according to the used protocol. The first type acquires the geolocation by deploying landmarks around the data center and then each landmark creates rings or circles around themselves using delay-to-distance positive correlation. The overlapping area of all landmarks is considered the location of the data center. The second principle is to assume that the landmark with the lowest response time and data center shares the same region. The hardware-based models use a hardware component as a root of trust. The hardware component provides a tamper-proof environment where the information required for proof of location can be stored. The hardware component later shares the details with a third party to verify its geolocation. 
\subsection{Problem Statement}
The geolocation of data on the cloud is a key concern for users. The data can be 
migrated to different hosts because of multiple reasons like load balancing, hardware upgradation, hardware failure, and energy saving. But moving the data to an unauthorized location could result in a violation of SLA and regulatory requirements. The data and all its replicas should be stored and processed in regions allowed by the user; but the flexibility of storing 
data in less expensive locations, without compromising the data sovereignty requirements, is also required. However, only secure and trustworthy hosts should be able to operate data. 

\subsection{Contribution}
The models that are based on Round Trip Time (RTT) to estimate the geolocation of data are error-prone and their limited scope to the specific type of storage demands something more 
advance. Therefore, we proposed a hardware-based model that uses TPM as a root of trust. First, we have presented a model that will attest the host based on their actual geolocation and allow the authorized host to store data in plaintext. Second, we have demonstrated the implementation of the proposed model using OpenStack. Lastly, we evaluated the level of security offered by our model and the performance of the proposed approach in a real cloud environment.

\subsection{Organization}
The rest of the paper is arranged as follows; In section 2, we have outlined the important terminologies required to comprehend this study. In Section 3, we performed a thorough 
literature review of the most relevant research articles. In Section 4,  we have discussed the 
proposed solution which is a hardware-based model to enforce the data geolocation policy. In 
In section 5, we have demonstrated the prototype implementation. Section 6 contains the results and analysis of the security and performance. In Section 7, we have concluded the study with our observation on the importance and the feasibility of our model.

\section{Background}
In this section, we presented the overview of some important terminologies that will help us to demonstrate our study in a good manner and will assist readers to understand the 
thesis. Section 2.1 highlights the general architecture or procedure of data storage in a cloud infrastructure. Section 2.2 differentiate data governance 
in cloud and non-cloud environments. Section 2.3 presents the importance of data geolocation in the cloud and legal aspects. Section 2.4 gives an overview of the Trusted Platform Module (TPM).
Section 2.5 explains the geolocation module which highlights the procedure of acquiring geolocation parameters.
\subsection{Data Storage in Cloud Environment}
Data storage is offered by cloud computing providers in a form of storage as a service.
The data uploaded by the customer is made available to authorized users over the
internet. The users are allowed to remotely upload and access the data via a web
interface. With the advancement in technology and more usage of wireless networks
and mobile devices, the usage of online storage has also increased \cite{Obrutsky2016}.
At the most basic level, the client will upload or access the data through a form of
read/write request. The request will be directed by the controller node to the respective
storage server. For all the read/write operations the user has to get an instance from
the cloud service provider with whom the storage will be mounted. When a user tries
to access his data, he uses that instance, and all of his data is represented to him in an
organized way. The cloud has a distributed nature and each physical host is shared
among multiple users courtesy of virtualization. In the case of data, it is divided into chunks and then stored on different hosts for sake of load balancing.

\subsection{Data Governance}
Data governance points to the practice of authority and control over the data. The organizations through data governance specify the roles and responsibilities such as in decision-making on the data of the organization. The key domains involved in data governance are 1) Data Governance Structure, which focuses on ensuring that the roles and responsibilities are introduced and assigned properly throughout different organizational levels. 2) Organizational, which includes the support of higher management in deploying the data governance policy. 3) Environmental points to the regulations and necessary legal compliances. 4) Technical, which addresses the technical aspects of data. 5) Measuring and Monitoring Tools to check if data is fulfilling all organizational objectives \cite{Abraham2019}. The aim of data governance policy is to improve 
the value of data and reduce the risk related to data. 


Data governance is different in the cloud and non-cloud environments \cite{Majid2016}. When customers outsource their data to the cloud, they no more have physical control over their data. The data governance in the cloud includes some additional parameters such as cloud deployment models which are public, private, hybrid, and community, and cloud service models such as Infrastructure as a Service, Platform as a Service, and Software as a Service. SLA, which is a contract between the customer and CSP comprising the details for security parameters, region, etc. Last, cloud actors are the individuals that will be handling data in a cloud environment\cite{Majid2019}.

\subsection{Data Geolocation and Legal Aspects}
According to the article \cite{Basu2018}, the geolocation of data on clouds is an issue that needs to be addressed. The CSP can move the data from one data center to another in the name of load balancing or hardware upgradation. This movement of data creates a regulatory compliance concern. The geolocation of data has an impact on the performance of applications. If the application is hosted in one region while the storage that is used by the application is in another region then it will cause an increase in latency and will result in a downfall in the efficiency of the application \cite{Mazumdar2019}.

To provide services in a certain region, the cloud service provider needs to be compliant with the regulations of that particular region. The GDPR is a requirement for offering services in the European region. According to the GDPR, the data of the nation should not be stored outside the authorized region or country. In the USA, USA PATRIOT Act states that the personal data of the citizens should be stored within the country. The authors in \cite{Lang2018} discussed that the security of data is directly affected by its geolocation on the cloud. This is because of local laws whose compliance is made necessary by the government. The more updated and strict laws mean more privacy and security of data.

\subsection{Trusted Platform Module}
 Trusted Platform Module (TPM) contains a tamper-proof module that is implemented by a chip\cite{Yan2020}. It can be used to resist software and hardware-level attacks. A TPM has a non-volatile memory, multiple Platform 
Configuration Registers (PCRs), and a cryptographic module to handle cryptographic operations. The objectives that can be attained by using TPM in the cloud are the remote attestation of the host’s software and hardware configuration to ensure trustworthiness, and integrity measurement by calculating and storing the hash of different components of the host, key management via a cryptographic module, trusted boot by extending the integrity measurement, binding and sealing keys for secure communication, identification of the host by unique TPM key, VM migration, and access right management \cite{Hosseinzadeh2020}.

\subsubsection{Platform Configuration Register (PCR)}
TPM PCR refers to the memory location inside the TPM. The hashes of host configurations including configured hardware and software can store in these PCRs. The size of these memory locations for storage purposes depends on the hashing technique associated with it. These hashes in multiple TPM PCRs collectively form a PCR policy. In our model PCR policy plays an important role. Whenever TPM generates Sealkey a PCR policy comprising the current PCR values is associated with keys. The policy makes sure that private sealkey can only be used if the current PCR values match with the one in the PCR policy attached to the sealkey pair.

\subsubsection{SealKey}

SealKey is a pair of asymmetric TPM keys that can be used for the encryption and decryption of data. The private sealkey will only be used if the current values in TPM PCR match with the values in the PCR policy associated with the Sealkey pair at the time of the creation of keys. As in our case, geolocation is important and will be extended to TPM PCR. Considering the geolocation factor, if the host's location is different from the location at the time of the creation of SealKey the TPM will not provide the SealKey to the host.
\subsection{Geolocation Module}

The geolocation by the host can be acquired at boot time. The geolocation can be obtained via two methods. First, use an internal Global Navigation Satellite System (GNSS) module that is embedded on the same board as TPM. There are processors available that contain internal GNSS modules. The geolocation parameters can be obtained and extended to TPM PCR. Second, external GNSS modules can be used to acquire geolocation. This will be a standalone device and can be placed separately or on the roof of the data center. The compute node can be connected to an external GNSS module via wire or it can get geolocation parameters from a centralized database at the boot time. The reverse geocoding \footnote{Reverse geocoding is a procedure to obtain the name of a city or country from longitude and latitude.} will be used on obtained geolocation constraints.

\section{Related Work}
The geolocation of data on public clouds is a major concern for users. Outsourcing the data to the public cloud causes the user to lose physical control over his/her data. There are multiple problems when outsourcing the data to an unknown location. The user might be working on a critical research problem that might be banned in the country where his/her data is stored by the CSP. The laws like GDPR, the USA PATRIOT act force the CSP to store the private data of citizens within the country or specific region defined by the users. The SLA governed the aspects of data geolocation and other security measurements to ensure the privacy and integrity of data. Above all, there is a need for a mechanism that ensures that data can only be processed and stored in plaintext in a user-specified location.

Different techniques have been considered to address the data sovereignty problem. As cloud computing is involved in everything, the concern about the geolocation of data on the cloud is increasing. In \cite{Sasubilli2021}, the importance of data geolocation in clouds is explained and different threats were discussed. Multiple techniques that are proposed to address the data sovereignty problem in the cloud were explained.

Over time, researchers proposed different solutions like landmark techniques, cloud framework techniques, and hardware-based techniques to address data sovereignty. But with the advancement in technology, new techniques are also acquired by the adversaries over time that put a question mark on the credibility of the proposed techniques. The researchers in \cite{Jaiswal2016} discussed different scenarios that prominent the importance of determining the geolocation of data when outsourced to the cloud. They proposed a solution to identify the location of data centers where the data of the user is being operated. The protocol selects a maximum number of three active Landmarks (LM) from a pool of active LMs. The list of zonal landmarks (i.e., sharing the same region) was initially created. Then the most suited three landmarks have been selected from the pool to share challenges with CSP. Initially, the Ping-RTT was used to determine the landmarks closer to the data center. The File Exchange - Round Trip Time (FE-RTT)is used to map the location coordinates. Landmarks will record the time required by CSP to transfer a required file to the landmark. The RTT during a file retrieval could be very large so CSP can easily download files from other data centers when a challenge arrives from the user. To overcome this problem, In \cite{Jia2020} the author presented an improved challenge-response mechanism. Instead of retrieving the whole file, they use metadata about blocks of data. The solution mainly improved the challenge creation process to make it harder for CSP to evade. The researchers used the random blocks and sectors from a random block of data to make the proof generation more challenging. The location of the Third Party Auditor (TPA) with the least response time is considered the location of the data center where data is stored. The solution has a major drawback, the CSP can increase the delay of all auditors except one i.e., forge the location. The RTT in public verification could be error-prone.

The RTT-based delay is not feasible. It could be affected by many factors. In \cite{Zhao2019} the researchers proposed One Way Delay (OWD) as an improved alternative to RTT. The OWD is measured while a packet is transferred by Host A to Server S and from Server S to Host B. They improved the triangulation model that uses rings instead of circles to determine a more accurate location. The overlapping area of the rings is considered the location of the data center. The Merkle Hash Tree (MHT) has been used to verify if the CSP has stored all blocks of data in one place. The leaf node of MHT contains signatures of blocks of data and the parent node contains the information about leaf nodes. Users keep the MHT as a receipt. To ensure that data is stored in the specified location by the CSP, the user will generate a challenge and share it with agents (at least three, efficiency depends on as many agents as possible) around the data center where data is expected. The agents will forward it to the CSP and wait for a response. The CSP must generate a signature of the enquired block with a random number (given in the challenge), and auxiliary information of its MHT and send it back to the agent. The agent will forward it to the user and the user will verify the proof and using the OWD will generate rings around distributed agents to determine the geolocation. The rings must be generated by the user and they should have overlapping areas (i.e., the location of the data center). There will be no overlapped area only if CSP tries to increase or decrease delay by certain numbers. The OWD makes it complex for the CSP to alter the delay because alteration on one delay affects at least two others. For higher efficiency, twelve auditors should be deployed around the data center. The researchers assumed that if we restrict 
the CSP to store the data only in user-defined regions then because of the cost factor the data will not be outsourced anywhere else. The delay-to-distance mapping models are dependent on response delays. The larger size of the challenge block will result in a larger response delay. Similarly, the researchers in \cite{eskandari2017dloc} proposed a model DLoc where the geolocation of the server where data is placed is calculated by generating rings according to the response time. The researchers neglected the network latency and possible manipulation of the time during the response generated by the insiders. The authors in \cite{Zhang2020} propose a more efficient protocol, Splitter. The Splitter can determine the location of data in less time. This was done by breaking up the challenge and proof generation procedure. The response of general operations (i.e., addition and multiplication) was measured and other exponentiation operations were ignored. The random forest algorithm and a better triangulation method (proposed by \cite{Zhao2019}) are used to identify the geolocation of data. The input for the challenge generation algorithm is different data block identifiers and a list of landmarks with a known location. The output of the algorithm will be sub-challenges (i.e., each sub-challenge for a unique block of data). The active landmarks will send the sub-challenges to CSP. The CSP will respond with sub-proofs. The sub-proofs to sub-challenges will be broadcasted to the landmarks and landmarks will generate timestamp RTT on receiving the proofs. A final proof (containing all sub-proofs) will be then shared with TPA to ensure the integrity of the proof.

The hardware-based solutions are widely used for the remote attestation of the host. The Trusted Platform Module (TPM) is a microprocessor embedded in the motherboard of the host that can be used for integrity measurement and remote attestation of the host. The TPM 2.0 is ISO compliant and is recently consented by several governments such as the USA, Japan, Russia, France, and many more \cite{Hosseinzadeh2020}.

The authenticity of the host operating the data is as important as the geolocation in which data can be processed. A mechanism that combines Proof of Data Possession (PDP) to ensure the geolocation of data and trusted computing to make sure that data is being processed by a trusted host has been proposed \cite{Fu2014}. The model uses TPM to generate the proofs for the challenges. The proof for the challenging data and platform state (including the software and hardware configuration) is stored in TPM PCRs. The user can attest to the host to identify whether its data is operated by a secure and authentic host or not. A delay-to-distance correlation has been used to determine the location of the data center. As per experiments by researchers, TPM version 1.2 can process a single proof within 9ms. It may affect the response when the number of challenges increases. The researchers in \cite{jiang2021reliablebox} proposed a combination of delay-based and trusted computing mechanisms known as ReliableBox. The model uses the combination of RTT and OWD to estimate the geolocation of the host by generating the rings. Symmetric cryptography is used to encrypt the key before outsourcing it to the cloud. The data is decrypted upon the retrieval request from the user. To ensure the integrity of the data, a hash of the file is calculated and preserved by the client, and on the retrieval the hash is compared against the original hash to ensure the originality of the data. To calculate the response against the upcoming challenges from the verifiers, Intel Software Guard Extensions(SGX) is used in ReliableBox. The Intel SGX provides a secure environment called an enclave. Access to the Intel SGX is secured by the processor. The Intel SGX is used to create responses to the challenges and a time token created by the Intel SGX is attached to the responses to prevent an adversary from manipulating the time during the operation. 

The state-of-the-art landmark-based can only test the geolocation of data and they cannot enforce the geolocation policy. The users sometimes need cold storage like AWS S3 Glacier\footnote{https://docs.aws.amazon.com/amazonglacier/latest/dev/introduction.html} which can be used for data archival and backups, the delay-to-distance mapping models cannot be applied in such cases.

The authors in \cite{Paladi2015}, proposed a hardware-based solution. The trusted state of the host in the data center has been calculated by combining the geolocation and platform state of the host. The daemon process D proposed to get the location. D will be using an external or internal Global Positioning System (GPS) device to get the current location of the host at boot time. The udev rules will direct the Daemon D process to acquire the location metrics from a specific address. According to the proposed solution, the host will get the keys to decrypt the data from a trusted third party if the host is in a location that is allowed by the user for processing the data. At the boot time of the platform, the TPM of the host will store the current location and platform state in its PCR. 
The stored information will be given to the Trusted Third Party (TTP) which is a third-party server for attestation. The TTP has a database of the trusted platform state of the hosts. The TTP will match the data obtained from the TPM of the host with the data in its database and then look for the location of the host and if it is authorized to access the data, the key will be sealed to the host. The location will tamper proof as it is calculated at the boot time and then extended to the TPM PCRs. 
Trusted Computing can ensure a secure and trustworthy platform for cloud users. The TPM is widely used by researchers in order to ensure Virtual Machine (VM) migration over a trusted host. In \cite{Aslam2021}, the researchers have proposed a protocol to securely migrate VM over trusted platforms i.e., with similar configurations. First, the user sets a policy and sends it to CSP. The VM is launched according to the parameters in the policy. Two important parameters were TAL and RefConf which represent a trust score and some cloud platform compliance standards respectively. The pair of TAL-RefConf will be checked before VM migration. The authors proposed a PTAA that will act as a trusted third party. The PTAA is a team of experts that can attest to the authenticity of the software stack running over a host. A host will receive Trust\_Token after being certified by the PTAA. The Trust\_Token has an expiry time. The hosts can attest to each other without communicating with PTAA until the Trust\_Token is valid. The Trust\_Token contains Trust\_Token id, Platform id, TAL, RefConf, timestamp, and a public key PK-BIND. The K-BIND is a TPM-based asymmetric key and is sealed to the platform configuration. The key in Trust\_Token will map the Trust\_Token to the host for which it was generated. After getting the Trust\_Token, the source platform (Ps) (from where VM is to be migrated) and destination platform (Pd) (to where VM will be migrated) will share their Trust\_Token. Both the Ps and Pd will verify each other’s Trust\_Token to check if the request is coming from an authentic source or if the destination host is reliable and according to the user-specified policies.

The authors of \cite{Cao2014} used a similar approach as \cite{Paladi2015}. The domain of focus was VM migration over trusted compute nodes. The trust was built on the basis of the platform state and geolocation of the computing node. The invention uses TPM as Secure Processing Unit (SPU) for the purpose of building a pool of trusted hosts on which VM can be migrated. The authors proposed to use several techniques to acquire the exact location of the host. The GNSS and cellular networks signal receiver can be used to identify the location of the host. The Inertial Measurement Unit (IMU) which includes Accelerometers, Gyroscopes, and Magnetometers that can be used to track the movement of the host. They can be placed in a trusted zone, the same as a GNSS receiver. If a movement is recorded by the IMU, the localization unit that contains the GNSS, cellular network receiver and IMU will again calculate the location of the host, and the third party will again remotely attest the host and check if its location is changed or not. If the location is changed and the host is in an unauthorized location, then it will not be counted in the trusted compute pool.

   \begin{table}[!ht]
  \caption{Summary of the Related Work}
  \label{t1}
\resizebox{0.5\textwidth}{!}{

\begin{tabular}{|p{1cm}|p{3cm}|p{3cm}|p{3cm}|p{4cm}|p{4cm}|}

    \hline
   \textbf{Ref\# \newline Year}  & \textbf{Technique} & \textbf{Domain}  & \textbf{Contribution} & \textbf{Limitation} \\
    \hline 
   
    \cite{eskandari2017dloc}, 2017 & Landmark Based & Data Sovereignty & A delay-to-distance mapping model. & Neglected manipulation in time and network delays. \\

    \hline 
   [\cite{Jia2020}], 2020 & Landmark Based & Data Sovereignty & Proof Replay	Improve the challenge-response mechanism by only using metadata. & Assuming that the data center and landmark with the lowest response delay share the same region. \\
     \hline 
    [\cite{Zhao2019}], 2019 & Landmark Based & Data Sovereignty & Proposed One Way Delay(OWD) method. & Assuming operation time on the server end. \\
     \hline 
    [\cite{Zhang2020}], 2020 & Landmark Based & Data Sovereignty & Proposed a challenge splitting method. & Using RTT, which is not feasible. \\

     \hline 
    [\cite{Fu2014}], 2015 & Combination of Hardware and Landmark-Based Technique & Data Sovereignty & Ensure trustworthiness of host using TPM & Generating proof inside the TPM with a delay-to-distance model. \\

    \hline 
    [\cite{jiang2021reliablebox}], 2022 & Combination of Landmark and Hardware-Based & Data Sovereignty & Data Integrity & Complex cryptographic operations. \\
     \hline 
    [\cite{Paladi2015}], 2014 & Hardware Based & Data Sovereignty & Protocol to report geolocation and platform state using TPM & Not evaluated in a real cloud environment. \\
     \hline 
    [\cite{Aslam2021}], 2021 & Hardware Based & VM Migration & A secure protocol to migrate VM between cloud platforms. & The geolocation aspect not covered. \\

    \hline 
    
    [\cite{Cao2014}], 2019 & Hardware Based & VM Migration & Multiple approaches to get location parameters. & Performing multiple operations on location parameters outside the SPU. \\
    
    \hline
    \end{tabular}
    }
\end{table}

\section{Proposed Model}
The Cloud Service Provider (CSP) provides the Storage-as-a-Service, which is acquired by many organizations and 
individual users. The regions where data should be stored can be defined in SLA but it needs to be enforced. Storage types like hot, warm, and cold are important and need to be considered while designing a mechanism that ensures the geolocation policies. By considering these demands, we proposed a model that will enforce the geolocation policies in the public cloud. The proposed model will cover all storage types. Additionally, it will verify the authenticity of the host where data will be stored by including the host configuration in the attestation phase. For demonstration, there are many platforms like OpenStack, CloudStack, and OwnCloud available that can be used to deploy public and private cloud models. In our proposed model, we will be using OpenStack\footnote{OpenStack is an open-source cloud framework that we use to design our proposed solution, therefore OpenStack components will be considered and used in the proposed solution.} to build a cloud infrastructure where the object storage service is offered via \textbf{Swift}. Swift is similar to the Amazon S3, Azure Blob storage \cite{bhonsle2023microsoft}, and Google Storage in Google Cloud \cite{kamal2020highlight}. Swift service is comprised of a \textbf{Proxy Server} that will route the end user requests to respective storage servers. A ring is a logical structure that contains the hash table which is used to map the name of files with their location. The proxy server uses the \textbf{ring} to identify the available storage servers where the data can be stored. Swift allows the creation of highly customized policies for data writing and retrieval. The policy can be used to add a geolocation-based filter while writing the data in Swift object. The storage server should have a TPM module enabled to store and provide the geolocation and platform configurations in order to get attested by the Third Party Attestation Server (TPAS). The end user will upload a
The description of the proposed model; corresponding steps are presented in Figure 1.
\begin{enumerate}
    \item The user will upload:
    \begin{itemize}
        \item Data: Data that will be encrypted and the key will be only shared between the client 
and Third-Party Attestation Server (TPAS).
\item List of authorized regions: A list of authorized regions where data should be stored 
in plaintext.
\item User signed token: A user signed token that will be used by Third Party Attestation 
Server to verify the list of authorized regions.
    \end{itemize}
    \item Proxy server will be using a policy that contains the list of the user-authorized regions for data storage.
\item The resource from the authorized region as per users' policy will be selected for the upcoming job.
\item The encrypted data on the selected resource will be written.
\item The encrypted data will be stored on the host from the authorized region.
\item The hosts from unauthorized regions will not be selected but can be used to write encrypted data replicas.
\item The list of hosts in authorized geolocation, and user-signed tokens will be shared with the TPAS.
\item Now the hosts have to share the TPM PCR values which contain their configurations 
and geolocation (described in Section 
4.2) with TPAS to get keys.
\item TPAS will not attest any host from unauthorized geolocation.
\end{enumerate}
As per the proposed model, the hosts being placed in the authorized regions and comprising TPM chips will be able to store and process data in plaintext.

\begin{figure}[H]
    \centering
    \includegraphics[width=0.5\textwidth]{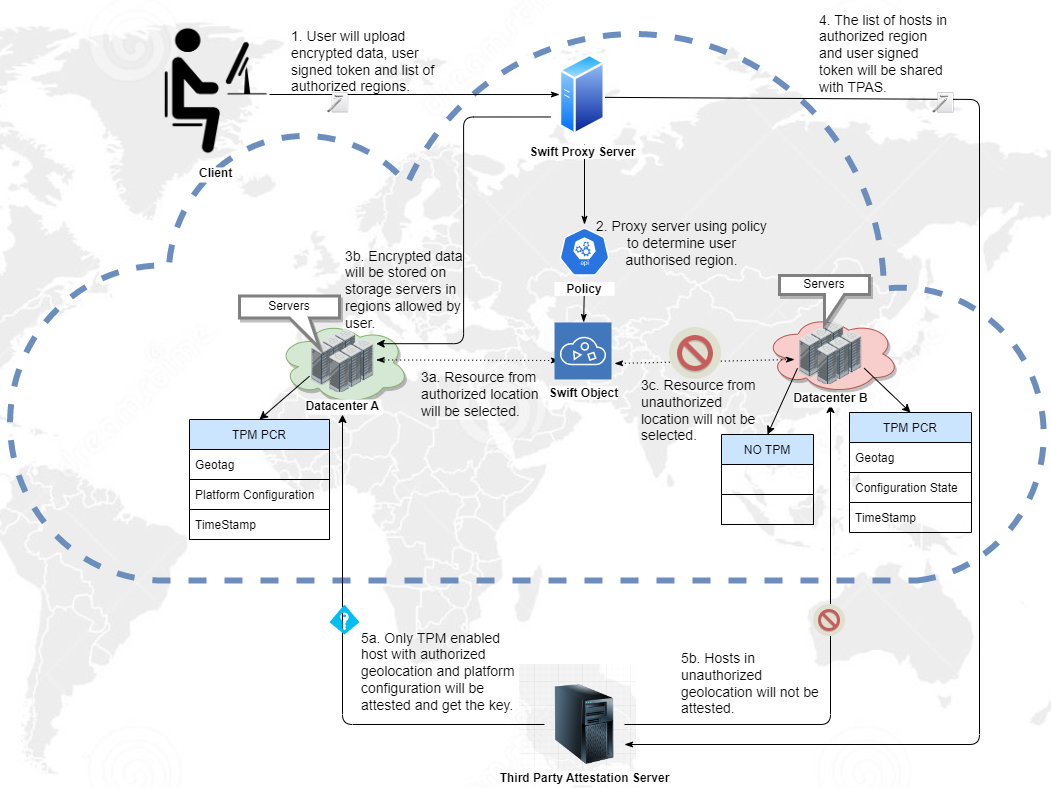}
    \caption{Proposed Model}
    \label{fig:1}
\end{figure}

\subsection{Key provisioning protocol}
The TPAS will ask for TPM PCRs in the challenge message. The TPM of the host will generate an asymmetric sealkey. The description of the sealkey is given in section 2.4.3. The TPM of the target host will generate a PCR quote containing platform configuration (PCR \{0-7\}) and geolocation (PCR \{15\}) using Attestation Identity Key (AIK).
The quote will be forwarded to TPAS. The TPAS will verify it against a good known value. Now the symmetric data encryption/decryption key \textbf{\textit{K}} will be encrypted with the target host TPM’s public sealkey. TPAS will share the key with the host. The \textit{\textbf{K}} will be sealed to the TPM in order to get it decrypted. The unsealing operation that will release the \textbf{\textit{K}} will only be possible if the host has authorized PCRs. PCR policy check will verify the PCR state. On a successful check, the key will be loaded into TPM memory and will be used to decrypt data.

\begin{figure}[H]
    \centering
    \includegraphics[width=0.5\textwidth]{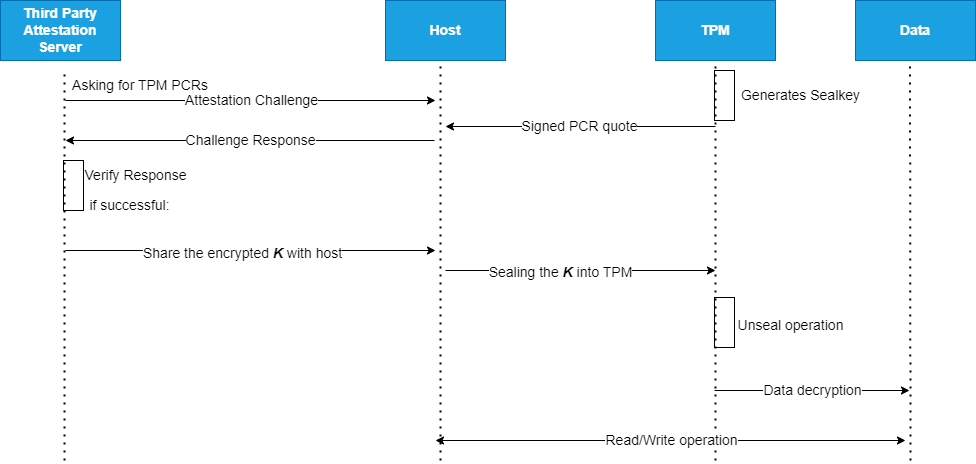}
    \caption{Key Provisioning Protocol}
    \label{fig:2}
\end{figure}

\subsection{Geolocation Protocol}
We created a daemon service that runs all the time on the host to get the geolocation constraints and extend them to TPM PCR \{15\}. The flow of the process can be seen in figure 3. The daemon is responsible for obtaining the geolocation, reverse geocoding the obtained parameters, and extending them to the TPM. GNSS module will provide the geolocation to the host. A daemon process (running all the time on the host) will obtain geolocation parameters from the GNSS device. Inside the daemon, there will be a reverse geocoding module to reverse geocode the obtained parameters. For the first time, the reverse geocoded location will be extended to TPM PCR \{15\} without any check. The periodic check module inside the daemon will periodically check the reverse geocoded geolocation of the host and if it found a different geolocation than the previously extended one, it will generate an event to extend the particular geolocation. The geolocation will be extended to TPM PCR \{15\}.
\begin{figure}[H]
    \centering
    \includegraphics[width=0.45\textwidth]{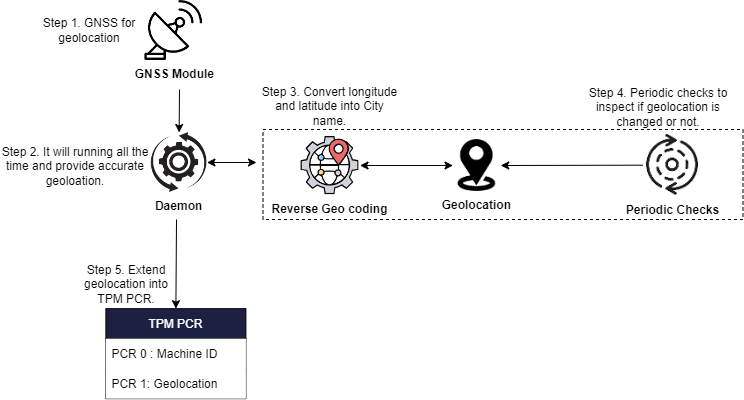}
    \caption{Geolocation Module}
    \label{fig:3}
\end{figure}
\section{Prototype Implementation}

In order to verify the sustainability of our proposed model, we used OpenStack, an open-source platform to deploy cloud infrastructure. A utility named Devstack\footnote{https://docs.openstack.org/devstack/latest/} is used to deploy the OpenStack. At the time of write-up, the Ubuntu 22.04LTS release is used in Oracle Virtualbox. We have deployed the latest version of OpenStack at the time. The DevStack comprises scripts that automate the installation and configuration of the multiple OpenStack services including nova, neutron, cinder, keystone, glance, placement, and horizon. We update the DevStack script to install and configure the Barbican key manager to store secrets and provide them to the services when needed. For the sake of demonstration, we used Virtual Trusted Platform Module (vTPM) which offers the same features as physical TPM. Oracle Virtualbox version 7 provides the support of vTPM version 1.2 and 2.0 using the libtpm and swtpm libraries. We also updated the /etc/nova/novacpu.conf file to create vTPM based compute flavors to enable vTPM support in the guest VM. 

In our model, geolocation has a key role in defining the host as authorized for data processing. For the demonstration, we use an android mobile as a GPS device. The ShareGPS android application is used for this purpose. We connected the mobile phone to our machine using Universal Serial Bus (USB) and made the daemon process read the geolocation from /media/device. On the host, we used gpsd \footnote{https://wiki.openstreetmap.org/wiki/Gpsd}, a utility to communicate with GPS devices, to get the geolocation coordinates. We built a service under /etc/systemd/system directory for the geolocation module. We updated the /etc/default/gpsd file and provide the interface address for external GNSS. Algorithm 1 is a high-level abstract of our geolocation module. The geolocation service requires initialized GPS module and gpsd on the host and will end up extending the geolocation in TPM PCR \{15\}. From the NMEA traffic, we captured the GPGGA message that includes the latitude and longitude and the number of satellites used to track the position of the host. The geolocation parameters in a GPGGA message that contains a minimum of three satellites are considered for further processing. The \textit{GPGGA.Lat} denotes the obtained latitude and \textit{GPGGA.Long} denotes the longitude parameter obtained from the GPGGA sentence. The obtained parameters are then reverse geocoded to get a more administrative unit. The \textit{Ga} denotes the hashed reverse geocoded geolocation. The \textit{Gdef} denotes the existing or default value of TPM PCR \{15\} and the combined hash of \textit{Ga} and \textit{Gdef} is denoted by \textit{Gexp} or expected geolocation whenever the host is rebooted. The \textit{Gcur} contains the existing or current value of TPM PCR \{15\}. Now, \textit{Gexp} and \textit{Gcur} are compared and if both are different the value \textit{Ga} which is the current geolocation will be extended to TPM PCR \{15\}. The ShareGPS application provides the geolocation parameter in NMEA format. 

\begin{algorithm}[H]
\caption{Geolocation Proctocol}\label{alg:custom top}

\begin{algorithmic}

\Require Initialized: GNSS Device, GPSD service
\Ensure Hash of geolocation extended to TPM PCR \{15\}
\State $START$
\While{$Number of Satellites \leq 3$}
    \State Get GPGGA message from GNSS Device
\EndWhile

\State $Lat \gets GPGGA.Lat$

\State $Long \gets GPGGA.Long$
\State $geolocation \gets REVERSEGEOCODE(Lat, Long)$
\State $Ga \gets SHA256(geolocation)$
\State $Gexp \gets Combined_HASH(Ga, Gdef)$
\State $Gcur \gets TPM2_READPCR(15)$

\If {$Gcur \neq Gexp$}

      $TPM2_PCREXTEND(Ga)$
\EndIf
\State $sleep(time)$
\State $END$
\end{algorithmic}

\end{algorithm}

In order to decrypt the data, the host needs to reply to the upcoming attestation challenge. Upon successful attestation, the data decryption key will be encrypted with a public sealkey and offered to the host. The host needs to generate Sealkey by extending the platform state to the PCR \{0-7\} and geolocation in PCR \{15\}. Algorithm 2 depicts the code used to generate the sealkey. The protocol will make sure that the data decryption key will be kept encrypted with the public SealKey and private SealKey will only be used as long as the host will have authorized values in its TPM PCR \{0-7\} and PCR \{15\} which will be platform state and geolocation of the host respectively.
\begin{algorithm}[H]
\caption{Sealkey Generation}
\label{alg:sealkey-pcr-policy}
\begin{algorithmic}

\Require PCR banks PCR \{0-7\} and PCR \{15\}
\Ensure Sealkey Pair

\State $START$
\State $TPM\textunderscore Startup()$
\State $ PCR\textunderscore Extend\{0\textunderscore 7\} : (Platform State)$
\State $ PCR\textunderscore Extend \{15\} : (Geolocation)$
\State $PCR \textunderscore SETPOLICY (PCR \{0-7\}, PCR\{15\})$
\State $ TPM \textunderscore Create(PCR \{0-7\}, PCR\{15\})$
\State $END$
\end{algorithmic}
\end{algorithm}

As we deployed the whole infrastructure on a single node, in order to simulate multiple storage disks we used loopback devices. The loopback device allows to access computer storage as a block file. Then we created mount points for these disks. Under the /etc/swift directory, we have to create configuration files for all of these disks to make them act as storage devices. Each of these configuration files contains paths to the respective disk. Now using the Swift Command Line Interface (CLI), we created a new ring that contains our loopback devices as random storage hosts in different regions. While adding the hosts in the ring, each host has metadata that represents the region. For sake of the experiment, we have added ten storage devices representing ten different regions. In order to upload the data in a specific region, the user has to give \textit{X-Object-Meta-Region:REGION\_NAME} value in the header while uploading the object. 

For data storage in a cloud environment, the user will first have to get authenticated with the keystone. Keystone offers the authentication service in OpenStack. Keystone will return the token that validates the authenticity of the user. The token associated with the write request will be routed to the proxy server which is the face of the Swift object storage service in OpenStack. Now the proxy server before writing the data to the Swift object will check the header tag that defines the user-preferred geolocation and place the data on that device that will match with the user-defined geolocation. Now, the data decryption key will be requested from the key management service using the key id. The key manager which is barbican in our case is using TPM as a key protector. The key is sealed into the TPM and will only be unsealed after satisfying the PCR policy. The unsealed key will be piped to the decryptor filter. Without storing the key on a disk it will be used to decrypt the encrypted object. The resultant decrypted object will be available for further processing on the Swift object and then the at-rest encryption as proposed in \cite{Daoud2022} can be implemented. Figure 4 depicts the working of our model.
\begin{figure}[H]
    \centering
    \includegraphics[width=0.5\textwidth]{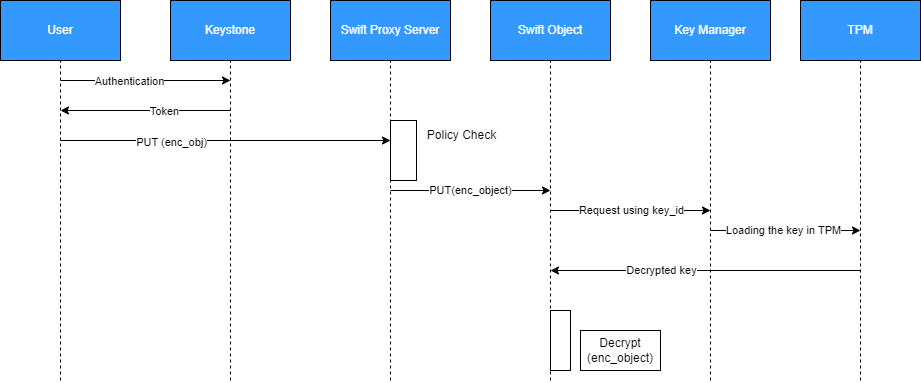}
    \caption{Flow diagram of PUT operation}
    \label{fig:4}
\end{figure}
\section{Results}
The technique for confining the geolocation of data in the cloud should be secure, efficient, and feasible to deploy. The approach should provide maximum efficiency without compromising the security of data. In this section, we have performed detailed security, performance analysis, and comparison with state-of-the-art solutions. The security analysis is performed to validate the model against possible adversarial threats. The performance of the model is estimated to check the efficiency of the proposed approach in the cloud. Last, we have compared our model with state-of-the-art solutions in terms of security and performance.
\subsection{Security Analysis}
Cloud service providers claim that they store users' data in a user-specified region. 
Improving the GNSS signal security is out of the scope of this work. The main focus of this study is to demonstrate the performance of the hardware-based model in the cloud environment. At the application level, to ensure that geolocation parameters are obtained from an authentic source, the udev \footnote{https://www.kernel.org/pub/linux/utils/kernel/hotplug/udev/udev.html} rule as explained by the authors in \cite{Paladi2015} can be used. The udev rule can bind the interface that will act as a source of geolocation parameters with the GNSS module. With the udev rule and daemon process running the geolocation of the host will be extended to TPM PCR\{15\} at the boot time. Extending this udev rule inside the Trusted Computing Base\footnote{https://www.ibm.com/docs/en/aix/7.2?topic=configuration-trusted-computing-base} will ensure its integrity. The periodic geolocation checks in our protocol are there to counter the term modular or containerized data centers that continuously move across different regions.

For data storage and usage in an authorized region, our proposed protocol relies on a remote attestation of the host on the basis of its geolocation and platform state. It will make sure that the data decryption key will only be made available to the host which is in the authorized region. The data decryption key encrypted with TPM public key will be decrypted with a private Sealkey and then get loaded into TPM memory. From the TPM memory, the key will be piped to the Swift pipeline and will be used to decrypt the data. The symmetric data encryption and decryption are never stored on disk in plaintext. A slight change in platform state or geolocation of the host in TPM PCRs would result in the unavailability of private SealKey which is required to decrypt the symmetric data encryption and decryption key. The different PCR values of three hosts A, B, and C respectively are mentioned in figure 5 which differentiates between authorized and unauthorized platforms for data storage. For the sake of clarity, we only represent the software state of the host denoted by \textit{Conf} in PCR lists. Host A comprising the authorized software state (\textit{Conf(A, B, C)}) and geolocation value in TPM PCR will get the symmetric data encryption and decryption key sealed into its TPM memory. The \textit{Host B}, containing an unauthorized software '\textit{H}' with the intention to dump the memory of the host and get the key in the plaintext will not be attested. Similarly, Host C comprising an authorized software state but being in an unauthorized jurisdiction will be not attested by the Third Party Attestation Server because of conflicting geolocation.

\begin{figure}[H]
    \centering
    \includegraphics[width=0.5\textwidth]{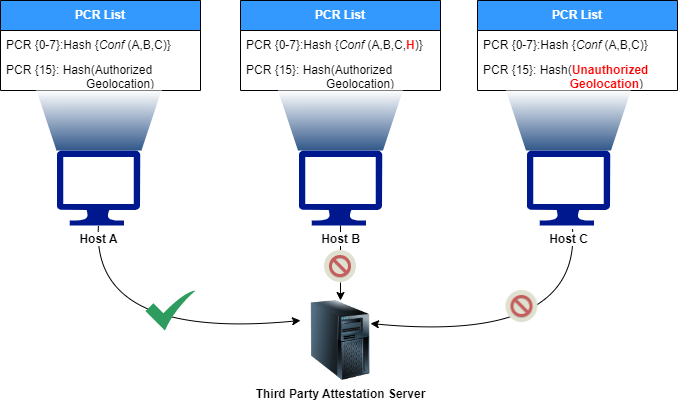}
    \caption{Authorised vs Unauthorised PCR states}
    \label{fig:my_label}
\end{figure}

\subsubsection{Adversary Model}
The adversary in our model will try to place the user's data in an unauthorized region or get a decrypted copy of data in an unauthorized region. Given the physical security of the host, in order to accomplish the objective, the adversary will be trying to place data on hosts similar to Host B and Host C mentioned in figure 5. First, if the adversary place data on Host B where \textit{H} configured software is used to get the memory dump in order to get the symmetric data encryption and decryption key in plaintext, the TPAS will not attest to such host even being placed in the authorized region. Because the PCR values shared with TPAS will also include the hashed list of software installed on the host. The TPAS will verify the software state of the host to be attested with a known good value that a host must comprise. If two values do not match the host will not be attested. The unauthorized software \textit{H} could be the result of both intended or unintended software installation. In both cases, the host will be considered unauthorized. Second, in the case of the host with PCR values similar to the Host C comprising \textit{Unauthorized geolocation} in TPM PCR \{15\} will not be attested. Given the physical security of the host, the adversary coming up with a platform state similar to Host B or Host C will not be able to access the data in plaintext. 

To communicate with the Central Processing Unit (CPU), TPM uses the serial peripheral interface (SPI) which is a communication interface. By default, there is no encryption method on SPI thereby, the data travels in plaintext on it. To intercept the communication between TPM and CPU comprising the data decryption key, one needs to intercept the communication on the SPI bus. For that adversary needs special equipment, enough knowledge of computer hardware, and physical access to the server. Physical access to the servers in data centers geographically located in different regions is normally not possible for an adversary. But in case an adversary gets physical access to the host with the administrator rights then it will be possible for him to extend his desired values in TPM PCRs. This can result in either denial of service in an authorized region by extending unauthorized geolocation in TPM PCR \{7\} of the host or by extending an authorized geolocation value in TPM of the host which is placed in an unauthorized region.

\subsubsection{Denial of Service Attack}
In our model, the denial of service attack can be launched on the host containing the user's data. The cloud computing architecture is robust enough to recover from failure. The replicas of data are stored on different hosts within the same or different data centers. In case of downtime of one availability zone or host, the user can be served with alternate options.
\subsubsection{CIA Triads}
For confidentiality, on the client side, the data is encrypted by the user with a symmetric but very unpredictable key. The user will share the key with TPAS by encrypting the plaintext symmetric key with the public key of TPAS to ensure that the key is only available in plaintext for the user and TPAS. After successful attestation of the host, the key will be sealed into the TPM memory using a public Sealkey. At runtime, to use the sealed key, the host must have authorized PCR values as described for Host A in figure 5. If by virtue of intention or unintentional, data or any of its replicas are stored on the host with PCR values similar to Host B or C it will not be attested by TPAS and encrypted data will be written on the disk.

The geolocation of the host on which users' data is being operated by the CSP should be according to the user-defined policy. The geolocation reporting protocol explained in section 4.2 assures the integrity of geolocation parameters. If there will be any unauthorized software configured on the host with the intention of faking the geolocation to get attested by TPAS, the proposed model is capable enough to detect it by evaluating the values in TPM PCR \{0-7\}.

For availability, with our protocol implemented users will still be able to choose multiple authorized regions. In case of hardware failure or upgradation, any other replica from the authorized region could be processable. Whenever data needs to be moved from one host to another it will first be encrypted by a symmetric key. The host where data is moved needs to be attested by the TPAS and after successful attestation, the host would be able to decrypt the data as per the procedure described in figure 4.

\subsection{Performance}
Our model adds a geolocation-based encryption layer on data to the existing methodology of cloud computing. The geolocation-based encryption layer is to control the movement of data in cloud infrastructure. In traditional cloud computing, the user uploads its data on the cloud the data is encrypted and then stored on disk. When the user makes a request for data it is first decrypted and then presented to the user. In our model, the user will outsource encrypted data to the cloud and at the cloud end, the data needs to be decrypted first to remove the geolocation-based encryption layer and to make it processable. After removing the encryption layer, the later process like at-rest encryption will be similar to the traditional cloud computing model. 

During the implementation, we observed a drop in the performance of the GPS module being placed in a basement lab. So, the usage of internal GPS is not recommended in such an environment. With the external GPS device, we get the geolocation parameters in ~30s from a cold start. The GPS module being placed at the same place, provides the geolocation parameters instantly during hot start. The ShareGPS application also provides the number of satellites used to generate a geolocation lock. The status \textit{3D} represents a minimum of three satellites used to obtain geolocation.

During normal write operations as per our model, the decryption of data is a major factor that increases the latency. For the sake of performance, in our model, we use a symmetric key for data encryption and decryption. After successful attestation of the host, the symmetric key sealed into the TPM of the host will be loaded into TPM memory and will be piped to decrypt the data during a write operation. As TPM is a slower device so the unseal operation on the basis of PCR \{0-7\} and PCR\{15\} to get the symmetric key is also an influencer on the performance of the model. The unseal operation is a one-time operation as long as the host attains authorized values in its PCR \{0-7\} and PCR \{15\}. As a result of the unsealing operation, the key is available to use and is kept in memory. In our experiments, the unseal operation takes 0.743s. To evaluate the efficiency of our model, we have used files with sizes between 1MB and 1000MB in write operations. We first uploaded the plaintext file assuming it as a normal write operation. There is no encryption or decryption involved in this operation. The write operations of plaintext file are denoted by \textit{black} line in figure 6. Second, we upload an encrypted file with a decryption key sealed into the configuration of the swift proxy server. The proxy server reads the decryption key, decrypts the data, and writes it on the disk. The operation is denoted by \textit{blue} line on the graph. Last, we implemented our protocol and upload encrypted files of size 1MB to 1000MB. This time the key is stored in the TPM. For every upcoming write operation, the key will be requested from the key manager barbican which is using TPM as key protector. The operation is denoted by the red line on the graph. In figure 6, it can be observed that there is a relatively minor overhead. The overhead percentage for files from the size between 1M and 1000M is ~0.50\%. The overhead is near constant and is because of key protection inside the TPM. Once, the key is loaded into TPM memory, it is requested for every write operation. Thus, this request for the key is only overhead in the process.

\begin{figure}[H]
    \includegraphics[width=0.5\textwidth]{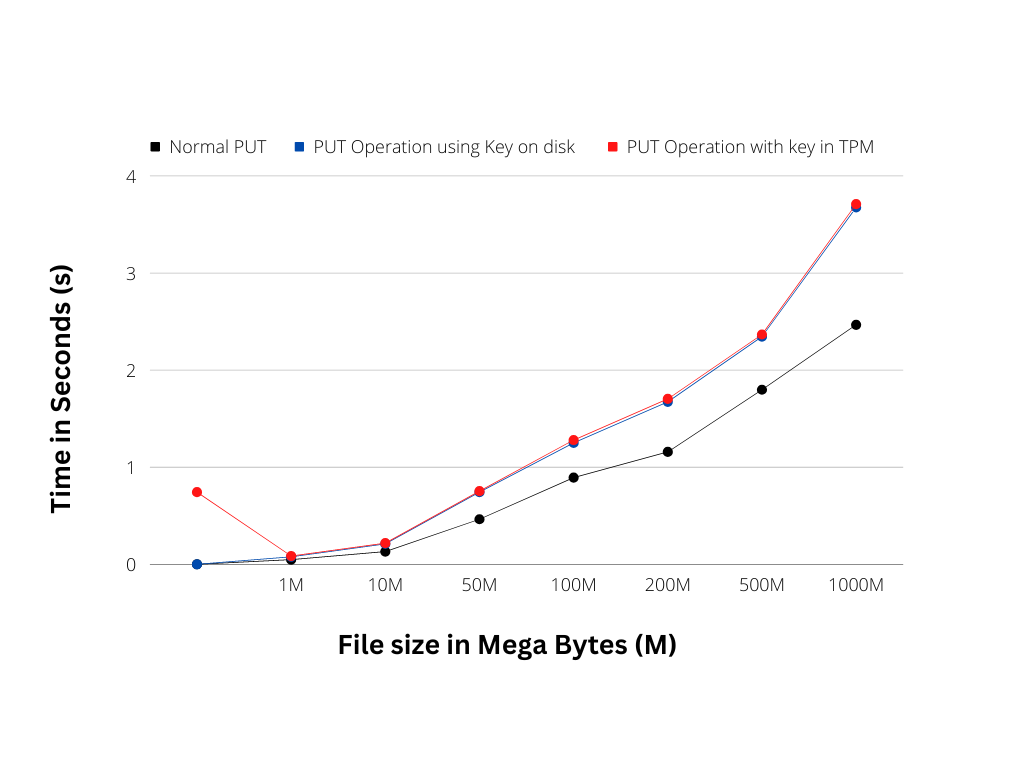}
    \caption{PUT operation}
    \label{fig:my_labe}
\end{figure}

Our proposed model is using existing technologies to enforce the user-defined geolocation policies while outsourcing its data to the cloud. We have used TPM as the root of trust for reporting the geolocation of the host. We have also used TPM as a key protector which ensures the confidentiality of symmetric data encryption and decryption key. The TPM device which is available on almost all PCs and servers and in the unavailability of TPM, the HSMs a standalone device to store passwords and secrets can be configured as a replacement for TPM. But the GNSS modules might not be available with every server or PC. To overcome this issue, we proposed to use an external GNSS module that will be reporting geolocation to a single database, and from that database, the geolocation can be read by our daemon geolocation service running on the host. To ensure the authenticity of the geolocation service the process is described in detail in section 4.2. The use of an external GNSS module placed on the roof of a building can result in the quick acquisition of geolocation parameters. As shown in experimental results, TPM unseals operation is a one-time operation as long as the host is kept up and running. The overhead due to the policy in Swift is minor and varies as per the number of allowed regions where data should be available in a processable form. This ensures the feasibility of our model.

\subsection{Comparison with state-of-the-art}
Compared with the state-of-the-art solutions, our model not just verifies the geolocation of data but it enforces the policy. The host located in regions other than user-authorized will never be attested. Further, the approach ensures the processing of data over a trusted host by combining the geolocation and platform configurations in the attestation phase. With secure access mechanisms being placed as suggested in \cite{xiong2019flexible}, the model will restrict the usability of data in user-specified regions only. Table 2, depicts the difference between the aforementioned solutions and our model. The first column focuses on data geolocation whether the approach is locating the data or not. In the next column, there is a check on the trustworthiness of the host that will be operating the data. Column number three and four represent the focus on hot and cold storage respectively. We have mentioned the two storage types specifically because most of the delay-distance mapping techniques are inadequate to cover the cold storage type.
\begin{table}[H]
  \caption{Comparison of Various Approaches}
  \label{t1}
\resizebox{0.5\textwidth}{!}{

\begin{tabular}{|p{3cm}|p{3cm}|p{3cm}|p{3cm}|p{4cm}|p{4cm}}

    \hline
   \textbf{Approaches}  & \textbf{Data Geolocation} & \textbf{Host Trustworthiness}  & \textbf{Hot Storage} & \textbf{Cold Storage} \\
    \hline 
   
    [\cite{eskandari2017dloc}] & \checkmark & $\times$ & \checkmark & $\times$ \\

    \hline 
   [\cite{Jia2020}] & \checkmark & $\times$ & \checkmark & $\times$ \\
     \hline 
    [\cite{Zhao2019}] & \checkmark & $\times$ & \checkmark & $\times$ \\
     \hline 
    [\cite{Zhang2020}] & \checkmark & $\times$ & \checkmark & $\times$ \\

     \hline 
    [\cite{Fu2014}] & \checkmark & \checkmark & \checkmark & $\times$ \\

    \hline 
    [\cite{jiang2021reliablebox}] & \checkmark & $\times$ & \checkmark & $\times$ \\
     \hline 
    [\cite{Jaiswal2016}] & \checkmark & $\times$ & \checkmark & $\times$ \\
     \hline 
    [13] & $\times$ & \checkmark & $\times$ & $\times$ \\

    \hline 
    
    Our Approach & \checkmark & \checkmark & \checkmark & \checkmark \\
    
    \hline
    \end{tabular}
    }
\end{table}
For the performance-based comparison, the delay-distance mapping models in \cite{Jaiswal2016, Jia2020,Zhao2019,Zhang2020, Fu2014} have completely different models as compared to ours. They only test the geolocation of the data by sending dedicated challenges for blocks of data. The ReliableBox proposed in \cite{jiang2021reliablebox} offers a secure approach as compared to the former. The model uses cryptographic operations to maintain the confidentiality of data, a delay-to-distance approach for geolocation estimation, and trusted computing for proof of data possession. The model provides secure access to the data but due to cryptographic operations during every data read/write request can add overhead to the performance. In contrast, our approach requires one-time attestation of the host as long as the host sustains the authentic configuration and geolocation. 

\section{Conclusion}

Using the literature review, we first validated the importance and requirement of the hardware-based model in order to enforce the geolocation policies in public clouds. The current techniques have a major drawback of dependency on round trip time to calculate the geolocation of data. This can only test the geolocation but cannot ensure it. This can lead to mission-critical data processing in unauthorized regions. In contrast with landmark-based approaches that are only feasible with hot storage, our model is feasible for all types of storage. We proposed a model that ensures the processing of data only in authorized regions. The protocol securely reports the platform configuration and geolocation of the host to the attestation server. The authorization policy that we defined during TPM key creation will make sure that unseal operation will only be possible in authorized configuration and region.  We evaluated the performance of our hardware-based model in a real cloud environment. From the experiments, the feasibility of our approach is highlighted. VM migration is not the primary objective of this study. But the model has the flexibility to expand and cover the VM migration as well. In the future, the approach has the potential to completely mitigate the need for TPAS by implementing more obfuscated access controls.



\bibliographystyle{elsarticle-num}
\bibliography{bibfile}



\end{document}